# Beam specific planning target volumes incorporating 4DCT for pencil beam scanning proton therapy of thoracic tumors

Liyong Lin[1]*, MingLei Kang[1]*, Sheng Huang[1], Rulon Mayer[2], Andrew Thomas[2], Timothy D Solberg[1], James E McDonough[1] and Charles B Simone II[1]

*Equally Contributing Authors

[1]Department of Radiation Oncology, University of Pennsylvania, Philadelphia PA

[2]Walter Reed National Military Medical Center, Bethesda MD

Contact information:

Liyong Lin, PhD, Assistant Professor

Department of Radiation Oncology, University of Pennsylvania

2326W TRC, PCAM, 3400 Civic Center Blvd

Philadelphia, PA 19014

Email: linl@uphs.upenn.edu

Phone: 215-615-5638

Fax: 215-349-5455

Minglei Kang, PhD, Medical physics resident

Department of Radiation Oncology, University of Pennsylvania

2326W TRC, PCAM, 3400 Civic Center Blvd

Philadelphia, PA 19014

Email: Minglei.Kang@uphs.upenn.edu

**Abstract**

The purpose of this study is to determine whether organ sparing and target coverage can be simultaneously maintained for pencil beam scanning (PBS) proton therapy treatment of thoracic tumors in the presence of motion, stopping power uncertainties and patient setup variations. Ten consecutive patients that were previously treated with proton therapy to 66.6/1.8 Gy (RBE) using double scattering (DS) were replanned with PBS. Minimum and maximum intensity images from 4DCT were used to introduce flexible smearing in the determination of the beam specific PTV (BSPTV). Datasets from eight 4DCT phases, using ±3% uncertainty in stopping power, and ±3 mm uncertainty in patient setup in each direction were used to create 8×12×10=960 PBS plans for the evaluation of ten patients. Plans were normalized to provide identical coverage between DS and PBS. The average lung V20, V5, and mean doses were reduced from 29.0%, 35.0%, and 16.4 Gy with DS to 24.6%, 30.6%, and 14.1 Gy with PBS, respectively. The average heart V30 and V45 were reduced from 10.4% and 7.5% in DS to 8.1% and 5.4% for PBS, respectively. Furthermore, the maximum spinal cord, esophagus and heart dose were decreased from 37.1 Gy, 71.7 Gy and 69.2 Gy with DS to 31.3 Gy, 67.9 Gy and 64.6 Gy with PBS. The conformity index (CI), homogeneity index (HI), and global maximal dose were improved from 3.2, 0.08, 77.4 Gy with DS to 2.8, 0.04 and 72.1 Gy with PBS. All differences are statistically significant, with p values <0.05, with the exception of the heart V45 (p= 0.146). PBS with BSPTV achieves better organ sparing and improves target coverage using a repainting method for the treatment of thoracic tumors. Incorporating motion-related uncertainties is essential in maintaining marginal coverage and homogenous dose of treatment targets.
.

**Introduction**

The variability and uncertainty in water equivalent thickness (WET) along the intended proton beam path within a patient can cause a target miss if not properly taken into account. Urie et al. proposed a correction strategy for the double scattering (DS) technique by minimizing compensator thickness over a smeared radius (i.e., the misaligned distance of the target), and extending the distal and proximal proton beam range by a percentage given by the uncertainty in converting CT Hounsfield units to stopping power. This "smearing" approach uses a perpendicular search radius, typically fixed at 5 mm, over the entire beam path to add treatment margin to the target along the beam path and was implemented in the treatment of thoracic tumors by Moyers et al. using the DS technique. In contrast to the implicit smearing embedded in the compensator design and direct manipulation of the distal and proximal ranges of DS beams, Park et al. introduced a beam specific PTV (BSPTV) initially proposed by Rietzel and Bert to explicitly include variation of water equivalent path length (WEPL) along each beam direction; in this manner the BSPTV can be used in pencil beam scanning (PBS) planning single field optimization (SFO). In the BSPTV method (Rietzel and Bert, and Park et al), distal and proximal water equivalent treatment margins (WETM) are converted to geometric treatment margins (GTM) that are calculated according to local tissue heterogeneity and added beyond the target to achieve a smearing effect in PBS, thus accounting for WET variations related to the fixed value of misalignment of tissue from motion and setup.

Neither Moyer's nor Park's approaches account for the fact that the magnitude of motion varies within different anatomic regions. The modeling of treatment margin due to organ motion can be further improved by using Maximum Intensity Projections (MIP) and Minimum Intensity Projections (MinIP) obtained from 4DCT (Underberg et al). Matney et al. and Flampouri et al. used 4DCT for evaluation of robustness and for construction of a BSPTV in DS treatment plans, respectively.

Respiratory motion patterns change among different segments of the thorax region (Keall et al.). The magnitude of motion at a patient's surface (i.e., at the beam entrance) can be significantly smaller than that of tumor motion (Liu et al. and Weiss et al.). Therefore, a fixed smearing value based on the magnitude of target motion would typically be too large for proximal regions along the beam path, and a BSPTV using a fixed smearing radius corresponding to tumor motion magnitude would be larger than needed. Multiple researchers have attempted to refine the BSPTV method using 4DCT for PBS treatment. Graeff et al. attempted to extend BSPTV to multiple field optimization (MFO) IMPT fields in GSI's in-house TPS. Knopf et al. calculated BSPTV as the union of multiple treatment targets (CTV + margins) over different phases of 4DCT using the PSI and NIRS in-house deformable registration algorithms.

PBS treatment planning for thoracic tumors can achieve better sparing of organs at risk (OAR) than IMRT techniques (Chang et al., Kang et al. and Zhang et al.). Despite the dosimetric advantages of PBS, DS and IMRT remain the methods of choice for treating complex thoracic tumors. This is primarily because commercial treatment planning systems (TPS) are unable to: (1) calculate proton spot delivery interplay with organ motion, which can lead to overdose and underdose within the treatment volume, and (2) determine appropriate treatment margins around the treatment target due to uncertainties of motion, stopping power and setup, which might result in partial anatomical miss or underdose of the target volume. Although Li et al. recently evaluated the systemic errors in 3D dose calculation that appear in the PBS treatment of thoracic tumors, there is no simultaneous comparisons of OAR sparing and plan robustness between PBS and DS treatments in this disease site (Wink et al.)

In this work, we describe the development of a patient field-specific BSPTV, that incorporates respiratory motion and stopping power uncertainties using 4DCT for PBS treatment of thoracic tumors. As reported by previous researchers (Graeff et al. and Knopf et al.), we calculated the change of water equivalence path length (WEPL) associated with organ motion and converted such WETM to GTM based on local tissue densities. In contrast to the conceptual demonstration by Graeff et al. for MFO application, we focus on SFO application in multiple thoracic patients. And unlike Knopf et al., who conceptualize the method of using the union of (multiple deformable targets + treatment margins) over different breathing phases in two different in-house TPS of PSI and NIRS, our treatment margin is applied to average CT images and the iCTV, which is the union of multiple CTV targets rigidly registered from eight phases in a commercially available TPS (Varian Eclipse, Palo Alto, CA). Compared to Knopf and Graeff's approach, we need only to calculate WETM and convert WETM to GTM once instead of eight or ten times (the # of breathing phases). This may lead to slightly larger target volumes than those of Knopf and Graeff, similar to the more generous margin calculation method proposed by Flampouri et al. for DS treatment. To illustrate the potential advantages of BSPTV using 4DCT for the treatment of thoracic tumors, we will demonstrate how such a method is equivalent to "flexible smearing" in DS and can generate a more appropriate treatment target volume for a typical patient.

A number of pioneering researchers have demonstrated the magnitude of overdose and underdose within the target volume caused by PBS delivery interplay with organ motion and how such deviation from the prescribed dose distribution is washed out with multiple fractions (Dowdell et al., Grassberger et al., Ritchter et al. and Li et al.). In this study we demonstrate such an interplay phenomenon for a representative thoracic patient.

The focus of this study is to establish a method of BSPTV calculation and organ interplay so that we can systemically report the dosimetric difference of PBS and DS treatment methods for ten thoracic tumor patients. Such methods and dosimetric studies can also serve as the platform and baseline for future improvements that can be brought by various motion mitigation strategies such as multiple beam paintings, gating and tracking techniques (Lomax et al.)

**Methods**

In conventional radiation therapy, the internal clinical target volume (iCTV) is delineated by a radiation oncologist and by the union of CTVs on the corresponding eight- or ten-phase images of a 4DCT on average CT images. To account for random and systematic errors of patient setup, an additional margin of 5-8 mm is used to expand an iCTV to a PTV.

Our methodology to calculate WETM and convert WETM to GTM is based on the phase and average images of 4DCT and implemented as a standalone MATLAB (MathWorks, MA) program. The generated BSPTV were imported into Eclipse TPS for subsequent optimization, thus providing better PBS treatment plan of thoracic tumors. Instead of using a "fixed smearing" distance to search for maximal WEPL change over a beam path, the maximal difference of WEPL between eight 4DCT phases and the average CT images was used as WETM along each beam path. The distal and proximal water equivalent margins ($DM_{we}$ and $PM_{we}$) due to organ motion (WETM) were calculated as the WEPL difference between the MIP, MinIP and average of the eight phases of the 4DCT, respectively (Equations 1 and 2). In this manner, different smearing radii along the beam path, i.e. "flexible smearing," were effectively used to calculate WETM. Such WETM were later implemented as additional voxels beyond treatment targets (i.e. GTM) based on the local Hounsfield units in the average CT images and the projected voxel size over the beam path, whose summation should equal to WEPL differences (Equations 3 and 4). The steps of conversion of WETM to GTM were omitted in the original smearing approaches by Urie et al. and Moyers et al. but implemented in recent works of Rietzel, Park, Flampouri, Graeff and Knopf. Figure 1-(a) shows an example of how a fixed smearing of 5 mm radius over average CT images overestimates a BSPTV compared to that of flexible smearing over 4DCT images. Two BSPTVs, generated using fixed smearing and flexible smearing, are shown for a gantry angle of 270 degree. Because the motion of the ribs is less than that of the target, a fixed smearing of 5 mm radius overestimates the distal and proximal BSPTV margins. For the regions close to the heart, however, the distal BSPTV margin from flexible smearing can be slightly larger than that from fixed smearing because cardiac motion can exceeds 5 mm. In addition, cardiac motion is not inadequately modeled by breathing-phase-based 4DCT images. Figure 1-(b) demonstrates how the beam specific range uncertainties are broken down into three aspects: internal motion (Motion), stopping power ratio (SPR) of medium-to-water versus Hounsfield units in CT images, and the accuracy of image guidance (Setup). The BSPTV calculation uses 3% of the proton range plus 1 mm related to CT/stopping powers and, for patient setup variation, 2 mm along AP/RL directions and 3 mm along the SI direction. As these three sources of uncertainties are most likely independent, a quadratic method was primarily used for the summation of three uncertainties; however, a linear summation was also used to ensure target coverage, providing there was no overlap with adjacent organs at risk. Figure 1-(c) shows the contributions to the BSPTV as a function of gantry angle. The contribution from Motion is larger than from CT/SPR or Setup and contributes over 90% of the Sum as determined by the following equations:

$$WETM_{distal}(motion) = \sum_i [WET(MIP)_i - WET(avg)_i] \qquad (1)$$

$$WETM_{proximal}(motion) = \sum_i [WET(avg)_i - WET(\min IP)_i] \qquad (2)$$

$$\sum_{j \subset GTM\ distal} [WET(avg)_j\ dl] = WETM_{distal}(motion, SPR, setup\ or\ sum) \qquad (3)$$

$$\sum_{j \subset GTM\ proximal} [WET(avg)_j\ dl] = WETM_{proximal}(motion, SPR, setup\ or\ sum) \qquad (4)$$

where *i* are the voxels along the beamlet from the patient surface to the proximal and distal surfaces of iCTV in Equation 1

and 2, respectively; *j* are the amount of voxels to be determined in the distal and proximal GTM regions in Equation 3 and 4, respectively; and where *l* and *dl* are voxels and projected voxel dimension along the beam path in average CT, respectively.

We further define two overlap volumes; one between the BSPTV and organs, and the other between the beam path outside of the BSPTV and organs. While the beam path overlap volume may receive less than the full prescription dose of 66.6 Gy, minimizing the overlapped volumes along with the BSPTV will allow for sparing of lung, heart and cord prior to the step of plan optimization. Figure 1-(d) shows how the Matlab code generates an optimized BSPTV and overlapped volume at an ideal gantry angle. In this figure, a gantry angle of 300 degree was selected by both Matlab and the dosimetrist since it corresponded to the long axis of the target and optimal BSPTV and overlap volumes. In contrast, a gantry angle of 180 degree chosen by the dosimetrist is suboptimal since the BSPTV and overlap volumes are both relatively large.

Motion, CT/SPR, Setup, Quadratic and Linear Summations BSPTVs were used in the optimization step of PBS treatment planning for each of the 10 patients evaluated. The BSPTVs were optimized with goals of covering 99% of the target volume with 99% of the prescription dose of 66.6 Gy in 1.8 Gy (RBE) per fraction, though the coverage of BSPTV could be selectively compromised if there was a critical OAR constraint. The OAR constraints within the optimizer were set according to the constraints used in the RTOG 1308 protocol, a phase III randomized trial comparing Photon Versus Proton chemoradiotherapy for inoperable stage II-IIIB NSCLC. Two fields were used in each PBS treatment plan, and each was individually optimized and summed at the last step without any further optimization. The treatment angles were kept the same in PBS plans as they were in DS plans unless the plan optimizer preferred other angles.

Any comparison of treatment plans depends on the characteristics of beam lines used. The spot and Bragg peak characteristics of the IBA pencil beam scanning beam line have been reported by Farr et al., Grevellot et al. and Lin et al., while the source size and Bragg peaks of the double scattering technique have been described by Slopsema et al. The spot sigma at most patient surfaces decrease from 7 to 3 mm for proton energies from 100 to 225 MeV. In order to treat tumors shallower than 75 mm, i.e. the proton range of 100 MeV beams, we utilize a U-shaped bolus (Both et al.) with a WET of 75 mm placed on the top and lateral sides of patient. Our proton couch also has a built-in bolus with a WET thickness of 65 (or 75 mm with an overlay) to enable treatment of superficial posterior tumors. The dose algorithm PCS13.0.24 was used for all PBS and DS calculations of Eclipse TPS. The commissioning of Eclipse has previously been reported by Zhu et al.

To facilitate plan comparison between PBS and DS techniques, several dosimetric indices of target iCTV were used in this work, as described by Flampouri et al.: the volume of 100% isodose line, $IDL_{100\%}$; the Heterogeneity index, HI=($D_{5\%}$- $D_{95\%})/D_{prescription}$; the Conformity index, $CI=IDL_{100\%}/V_{ctv}$; Coverage quality, $CQ=D_{min}/D_{prescription}$; and Integral dose, $ID=V_{body}\times D_{avg}$.

Lomax et al. displayed uncertainty bands on the DVH curve to designate the uncertainties due to stopping power and patient setup. The uncertainty band method was also implemented by Lin et al. and Vargus et al. for OAR and target coverage for patient setup variation. In this work, a total of 8×12×10=960 plans were evaluated for ten patients, comprising permutations of eight 4DCT phases, ±3% stopping power, and ±3 mm x/y/z in patient setup. Percentiles of 25%-75% of all the permutations from the average DVH of OARs and iCTV were calculated for a representative patient and all ten patients (Figure 2 and Figure 3). The difference between PBS and DS was also calculated and displayed.

Ten consecutive patients treated with DS for stage III locally advanced non-small cell lung cancer were selected for the evaluation of PBS plans. All patients had mediastinal nodal metastases with various primary tumor locations. The comparisons of DVH criteria were evaluated using a paired t-test for statistical significance.

The method we used to evaluate beam interplay is similar to Zou et al. for conventional lung SBRT applications and other researchers for PBS applications (Grassberger et al., Dowdell et al., Graeff et al., Knopf et al. and Li et al.). A representative patient is shown in Figure 4, with a breathing period of ~3.5 seconds and delivery durations of ~60 seconds and ~46 seconds for the two beams. Switching time between energy layers, slew duration between spots and spot duration of the same energy layer were extracted from the beam delivery log files. For different treatment fractions or beam paintings within the same fraction, the beams start from a random position in the 3.5-second breathing period. PBS spots were therefore grouped into the eight different breathing phases. A single treatment plan was therefore split into eight plans outside of Eclipse using our in-house Matlab programs and a Dicom editor provided by Ion Beam Applications (Louvain-La-Neuve, Belgium). These plans were subsequently reimported into Eclipse TPS to calculate the dose distribution of each CT phase. The eight dose distributions were deformed and registered to the exhale-50 phase and summed in MIM Maestro (MIM Software Inc., Cleveland, OH).

**Results**

Table 1-(a) shows the magnitude of motion at three locations along the beam path for the ten consecutive patients evaluated. Tumor motion was larger than that of the ribs or at the patients' surface (beam entrance), and it was largest in the SI direction (~8 mm) followed by AP (~4 mm) and RL (~2 mm) directions. In contrast, the average motion of the ribs and patient surface at the beam entrance were less than 1.5 mm and 1 mm, respectively. Out of ten patients, all of whom were treated with two-field plans, six plans were treated using posterior and posterior oblique beams only, two plans involve a lateral beam, and the other two plans included an anterior beam. With supine positioning, there is minimal motion of ribs and posterior beam entrance. In contrast, when lateral or anterior beams were used, the motion was larger. Table 1-(b) shows BSPTV as a function of source of uncertainty. In the summed BSPTV, motion-induced range uncertainties were the largest contribution to treatment margins.

Figure 2 shows a representative patient treated with DS who would have benefited from a PBS technique. PBS spared both the high and low dose regions and allows reduced doses to the lung, heart, cord and esophagus compared with DS technique, while still maintaining 99% dose coverage of 99% of the iCTV. Furthermore, dose inhomogeneity and hotspots within the target were minimized, whereas target coverage was maintained for both PBS and DS (within ~2% dose) when 4DCT, stopping power and patient setup uncertainties were incorporated.

Figure 3-(a) shows the average, 25$^{th}$ and 75$^{th}$ percentile DVH bars for lung, heart, cord and esophagus of all ten patients for PBS (red) and DS (blue). P-values >0.05 suggest that the PBS advantage observed may not be statistically representative of each individual patient.

PBS plans were superior to DS in all cases. The greatest improvements with PBS was primarily within the low (V5) to moderate (V10 and V20) dose regions, with a reduction in mean lung dose (MLD) by 2.3 Gy from 16.4 Gy with DS to 14.1 Gy, and heart V30 from 10.4% to 8.1% with PBS (Table 2). There is a large variation in heart dose among the ten patients. Because one patient had almost no dose to the heart, the maximal heart dose average over ten patients fell outside of the 25 and 75 percentiles for both PBS and DS treatments. Excluding this patient, PBS had a maximal heart dose of 70.9 Gy averaged over nine patients, with 25 and 75 percentiles from 70.3 to 71.3 Gy, while DS had a maximal heart dose of 74.1 Gy averaged over nine patients, with 25 and 75 percentiles from 70.8 to 75.1 Gy. The reduction in heart V45, was not statistically significant (p=0.146).

In keeping with the dose constraints from RTOG 1308, only the maximum dose was constrained for esophagus. Despite this, PBS still resulted in a reduction in the average volume of esophagus irradiated. A p value of 0.049 indicates that the $D_{max}$ for PBS can be larger than for DS for some patients, although this occurs only when $D_{max}$ is well below 74 Gy. Because one patient had a maximal esophagus dose of 48 Gy in PBS treatment, the average over 10 patients fell outside of the 25 and 75 percentiles. Excluding this patient, PBS had a maximal esophagus dose of 70.0 Gy averaged over nine patients, with 25 and 75 percentiles from 69.5 to 71.2 Gy, while DS had a maximal esophagus dose of 72.6 Gy, with 25 and 75 percentiles from 70.5 to 73.5 Gy. The reduction in maximum heart and esophageal dose for certain patients from 72 Gy to 68 Gy (Table 2 and Figure 3) suggests that additional patients would benefit from PBS.

Table 2 shows that the integral dose (ID) is reduced by nearly 17% in PBS compared with DS. Additionally, the irradiated volume of the prescription dose (CI) is only reduced by 10%. The target coverage quality CQ is identical between PBS and DS, as 99% of the iCTV was required to receive the prescription dose with both techniques. In contrast, PBS results in a significantly more homogenous dose, with a homogeneity index HI of 0.04 compared with 0.08 for DS. Except for the marginal advantage of CI ($p = 0.056$), all the differences above were highly statically significant, with $p < 0.01$.

The robustness of iCTV coverage is shown in Fig. 3-(b) for all ten patients. For all PBS and DS plans, the iCTV receives no less than 97% of the prescription dose when all permutations of motion, CT/stopping power and patient setup uncertainties are included. Despite many clinical fears over overdose/underdose within the treatment target and target miss, it seems that PBS can potentially provide robust target coverage comparable or superior to that of DS when uncertainties are properly accounted for in the BSPTV with a repainting method to be discussed.

**Discussion**

Flampouri et al. proposed two methods to improve BSPTV in DS over the approach of Park et al by incorporating respiratory motion using two methods. The first uses MIP and MinIP data from 4DCT to determine the distal and proximal margins, respectively. In the second, the 4DCT phase that generates the largest margins is selected for each beamlet (beam path) to the target. The beamlet method results in a smaller BSPTV than the MIP/MinIP method. Unlike DS, however, in which all energy layers are delivered within 0.1 seconds, PBS delivers different energy layers sequentially over minutes. Thus delivery of the distal and proximal layers are out of phase and, in contrast with DS, more generous margins must be determined from MIP/MinIP instead of the worst 4DCT phase in PBS. We demonstrated that the contribution from motion is the dominant effect over stopping power and setup uncertainties to the treatment margins in PBS and concluded that "flexible smearing" based on MIP/MinIP is crucial to the success of PBS plan optimization. We further included the overlapped volumes of beam path with organs. Unlike DS, which always over treats proximal organs, sparing of such OARs by PBS beam can be substantial.

Our finding that the iCTV is more homogenously covered in PBS than in DS does not account for organ interplay with beam delivery in PBS. Accounting for interplay, the target dose distribution in each of 37 fractions of PBS is more heterogeneous than the planned distribution. The breathing period of ~3.5 second is comparable with the switching time between energy layers (1-2 seconds for our PBS system) and also comparable with the time required to paint each energy layer (with 3-5 ms per spot and 2-4 ms slew duration between adjacent spots with a total of ~6000 spots over ~20 energy layers). Our study indicates that an energy layer can be delivered in 1-4 out of 8 different breathing CT phases of different breathing cycles. As our spot size is relatively small, more fractions might be needed to achieve total dose homogeneity compared with centers employing larger spot sizes. Figure 4 shows the interplay on an axial slice: underdose (95%) and overdose (110%) are present for a single fraction without any motion mitigation. As a worst case scenario, dose heterogeneities from 90-120% can occur in a single fraction. However, such a magnitude of overdose and underdose within the target will be reduced when more treatment fractions are used. With 4 fractions, the extent of such overdose drops to 105%, with no 95% underdose, and with 8 fractions the 105% overdose volume decreases even further. The summated distribution from the deformable registered dose distributions of four fractions accounting interplay has a slightly larger area of 105% isodose and slightly smaller minimal dose than the planned dose distribution. With eight fractions, there was no observable difference for the DVH of the iCTV that accounted for interplay and the original treatment plan without interplay. As multiple beam paintings of 4 are clinically achievable, we believe that the dosimetric advantage of PBS treatment in this patient will be able to be maintained despite the interplay phenomenon. These observations are consistent with that reported by many researchers (Rietzel et al., Grassberger et al., Dowdell et al., Li et al. and Ritcher et al.). Because the breathing period and beam duration can vary from patient to patient, detailed assessment of the interplay effect for multiple patients and multiple motion mitigation strategies is the subject of current investigation within our group. Interplay is not an issue in DS, since all energy layers are painted every 0.1s.

Any plan comparison between PBS and DS techniques depends on the spot size of the PBS system, the placement of bolus/range shifter, the source size of the DS system, and the choice of prescription dose. The in-air proton spot size of our beam lines is smaller than those of the Hitachi system reported by Gillin et al. Furthermore, because the built-in posterior bolus and U-shaped anterior and lateral bolus are closer to the patient surface than range shifters mounted on the PBS nozzle, the spot size incident on the patient is smaller than systems that use nozzle-mounted range shifters. Therefore, conclusions drawn from this study may not be applicable to other beam lines.

**Conclusion**

The use of a beam-specific PTV based on respiratory motion along the beam path derived from 4DCT and incorporating CT/stopping power and patient setup uncertainties can achieve better organ sparing with PBS compared to DS.

Additionally, our study indicates that PBS planning based on BSPTV method can potentially achieve both better target homogeneity and robustness of target coverage using a repainting delivery method. Finally, beam angle optimization provides a reliable method to minimize doses to the critical organs.

**Acknowledgments**

The authors would like to thank Olivier De Wilde from Research and Development Division of Ion Beam Applications to extract the delivery log of the PBS treatment plan and Kevin Teo from University of Pennsylvania for deformable image registration using MIM software, respectively. This work was supported by the US Army Medical Research and Materiel Command under Contract Agreement No. DAMD17-W81XWH-07-2-0121 and W81XWH-09-2-0174. Opinions, interpretations, conclusions and recommendations are those of the author and are not necessarily endorsed by the US Army.

**Reference**

1. Urie M, Goitein M, Wagner M. Compensating for heterogeneities in proton radiation therapy. *Plasma Sources Sci Technol* 1984; **29**:553–566.

2. Moyers MF, Miller DW, Bush DA et al. Methodologies and tools for proton beam design for lung tumors. Int J Radiat Oncol Biol Phys 2001; **49**:1429–1438.

3. Park P C, Zhu X R, Lee AK et al. A beam-specific planning target volume (PTV) design for proton therapy to account for set up and range uncertainties. *Int J Radiat Oncol Biol Phys* 2012*;* **82:**329–36.

4. Underberg RWM et al. "Use of maximum intensity projections (MIP) for target volume generation in 4DCT scans for lung cancer." *Int J Radiat Oncol Biol Phys* 2005; **63**:253-260.

5. Rietzel E and Bert C "Respiratory motion management in particle therapy" *Med. Phys.* 2010; **37:**449–60.

6. Matney J, Park P C, Bluett J et al. "Effects of respiratory motion on passively scattered proton therapy versus intensity modulated photon therapy for stage III lung cancer: are proton plans more sensitive to breathing motion?"*Int. J. Radiat. Oncol. Biol. Phys.* 2013; **87**:576–82.

7. Flampouri S, Hoppe BS, Slopsema RL et al. "Beam-specific planning volumes for scattered-proton lung radiotherapy." *Phys Med Biol* 2014*;* **59**:4549-58.

8. Chang, JY, Zhang X, Wang X et al. "Significant reduction of normal tissue dose by proton radiotherapy compared with three-dimensional conformal or intensity-modulated radiation therapy in Stage I or Stage III non–small-cell lung cancer." *Int J Radiat Oncol Biol Phys* 2006; 65: 1087-1096.

9. Kang Y, Zhang X, Chang JY et al. "4D proton treatment planning strategy for mobile lung tumors." *Int J Radiat Oncol Biol* 2007; 67:906-914.

10. Zhang X, Li Y, Pan X et al. "Intensity-modulated proton therapy reduces the dose to normal tissue compared with intensity-modulated radiation therapy or passive scattering proton therapy and enables individualized radical radiotherapy for extensive stage IIIB non-small-cell lung cancer: a virtual clinical study." *Int J Radiat Oncol Biol Phys* 2010; **77**:357-366.

11. Dowdell S, Grassberger C, Sharp GC et al. "Interplay effects in proton scanning for lung: a 4D Monte Carlo study assessing the impact of tumor and beam delivery parameters." *Phys Med Biol* 2012*;* **58***:* 4137.

12. Grassberger C, Dowdell S, Lomax A et al. "Motion interplay as a function of patient parameters and spot size in spot scanning proton therapy for lung cancer." *Int J Radiat Oncol Biol Phys* 2013; **86**: 380-386.

13. Richter, DA, Schwarzkopf J, Trautmann M et al. "Upgrade and benchmarking of a 4D treatment planning system for scanned ion beam therapy."*Med Phys* 2014; **40**:051722.

14. Li Y, Kardar L, Li X et al. "On the interplay effects with proton scanning beams in stage III lung cancer." *Med Phys* 2014; **41**:021721.

15. Lomax AJ. "Lomax, A. J. "Intensity modulated proton therapy and its sensitivity to treatment uncertainties 1: the potential effects of calculational uncertainties." *Phys Med Biol* 2008; **53**: 1027.

16. Lomax AJ. "Intensity modulated proton therapy and its sensitivity to treatment uncertainties 2: the potential effects of inter-fraction and inter-field motions." *Phys Med Biol* 2008; **53**: 1043.

17. Wink K, Roelf E, Solberg T et al. "Particle therapy for non-small cell lung tumors: where do we stand? A systematic review of the literature" Frontiers in Radiation Oncology 2014; **4**: 00292. DOI: 10.3389.

18. Lin L, Vargus C, Hsi W et al. "Dosimetric uncertainty for prostate cancer proton therapy." *Med Phys* 2008; **35**:4800-4807.

19. Vargus C, Wagner M, Mahajan C et al. "Proton Therapy Coverage for Prostate Cancer Treatment." *Int J Radiat Oncol Biol Phys* 2008; **70**:1492-1501.

20. Keall P, Mageras GS, Balter JM et al. "The management of respiratory motion in radiation oncology of AAPM Task group 76." *Med Phys* 2006; **33**:3874-3900.

21. Liu HH, Balter P, Tutt T et al. "Assessing respiration-induced tumor motion and internal target volume using four-dimensional computed tomography for radiotherapy of lung cancer." *Int J Radiat Oncol Biol Phys* 2007; **68**: 531-540.

22. Weiss E, Wijesooriya K, Dill SV and Keall PJ. "Tumor and normal tissue motion in the thorax during respiration: Analysis of volumetric and positional variations using 4D CT." *Int J Radiat Oncol Biol Phys* 2007; **67**: 296-307.

23. Graeff C, Durante M and Bert C. "Motion mitigation in intensity modulated particle therapy by internal target volumes covering range changes" Med Phys 2012; **39**: 6005-6013.

24. Knopf AC, Boye D, Lomax A and Mori S "Adequate margin definition for scanned particle therapy in the incidence of intrafractional motion" *Phys. Med. Biol.* 2013; **58:** 6079–94

25. Li H, Liu W, Park P et al, "Evaluation of the systematic error in using 3D dose calculation in scanning beam proton therapy for lung cancer" JACMP 2014; **15**: 47-56.

26. Farr J et al "Fundamental radiological and geometrical performance of two types of proton beam modulated discrete scanning systems." *Med Phys* 2014; **40**:2101-8.

27. Lin L, Ainsley CG, McDonough JE. "Experimental characterization of two-dimensional pencil beam scanning proton spot profiles" *Phys Med Biol* 2013; **58**: 6193-6204.

28. Lin L, Kang M, Solberg T et al. "Experimental validated pencil beam scanning source in TOPAS." *Phys Med Biol* 2014; **59**:6859-73.

29. Grevillot L, Frisson T, Zahra N et al "Optimization of GEANT4 settings for proton pencil beam scanning simulations using GATE" *Nucl. Instrum. Methods Phys. Res.* B 2014; **268:**3295–305.

30. Gillin M, Sahoo N, Bues M al et al "Commissioning of the discrete spot scanning proton beam delivery system at the University of Texas M.D. Anderson Cancer Center, Proton Therapy Center, Houston" *Med Phys* 2009; **37**:154-63

31. Slopsema R, Lin L, Flampouri S et al. "Golden beam data for a double-scattering system" *Med Phys* 2014; **41**:091710.

32. Both S, Shen J, Maura K et al. "Development and clinical implementation of a universal bolus to maintain spot size during delivery of base of skull pencil beam scanning proton therapy." *Int J Radiat Oncol Biol Phys* 2014; **90**:79-84.

33. Zhu R, Poneisch F, Lii M et al "Commissioning dose computation models for spot scanning proton beams in water for a commercially available treatment planning system" *Med Phys* 2013; **44**:1723-37.

34. Zou W, Yin L and Shen J et al. "Dynamic simulation of motion effects in IMAT lung SBRT." *Radiation Oncology* 2014; **9**: 225.

Figure 1: (a) Two axial slices of BSPTVs were generated based on flexible smearing over 4DCT images versus 5 mm fixed smearing over average CT images at a gantry angle of 270 degree. The red, green and yellow contours are iCTV, BSPTVs from flexible smearing and fixed smearing, respectively; (b) A single axial slice showing BSPTV for another gantry angle of 300 degree with components originating from uncertainties in patient setup (Setup), CT/stopping power ratio conversion (SPR), 4DCT motion (Motion) and their linear and quadratic summations; (c) BSPTV volumes as a function of gantry angle with the quadratic ally and linearly summated BSPTV (QuaSum, LinSum); (d) Overlapped volumes of the BSPTV and beam path with lung, heart and cord are plotted as a function of gantry angle.

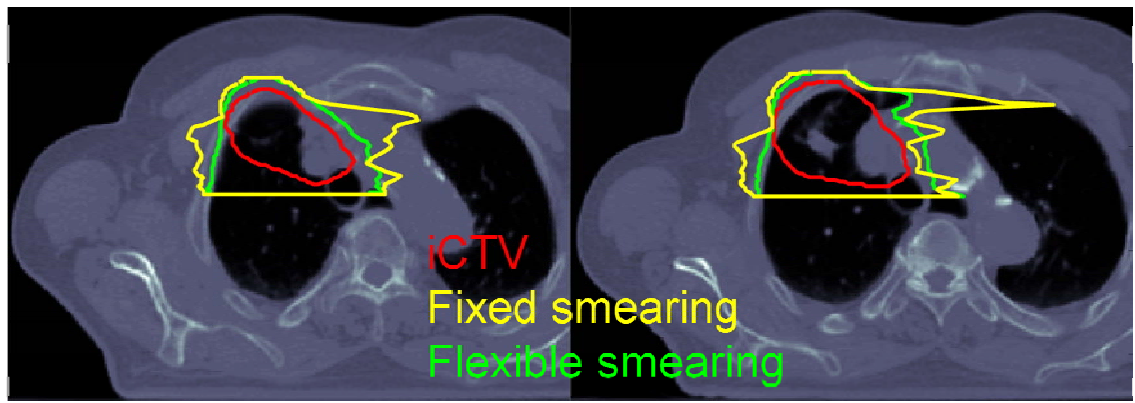

(a)

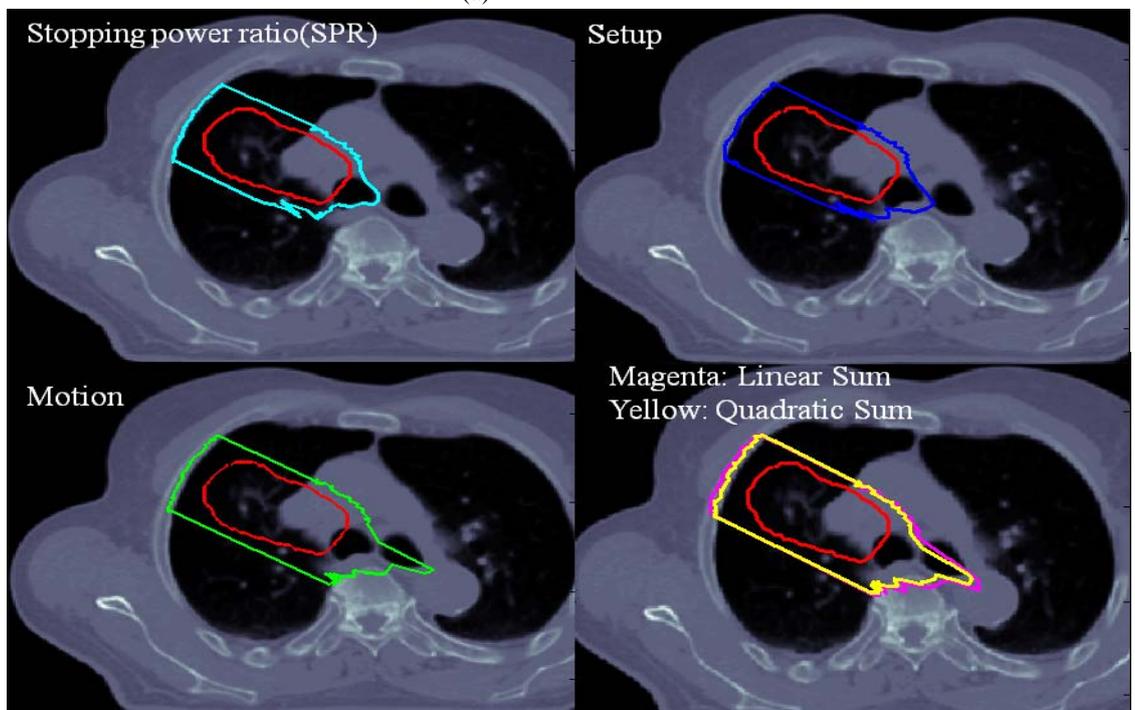

(b)

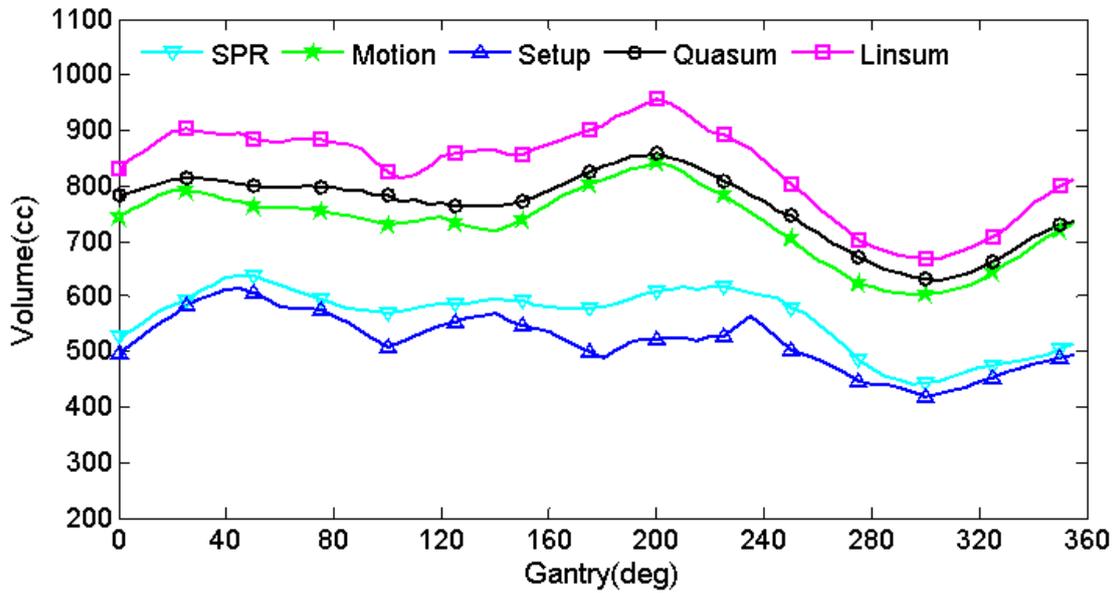

(c)

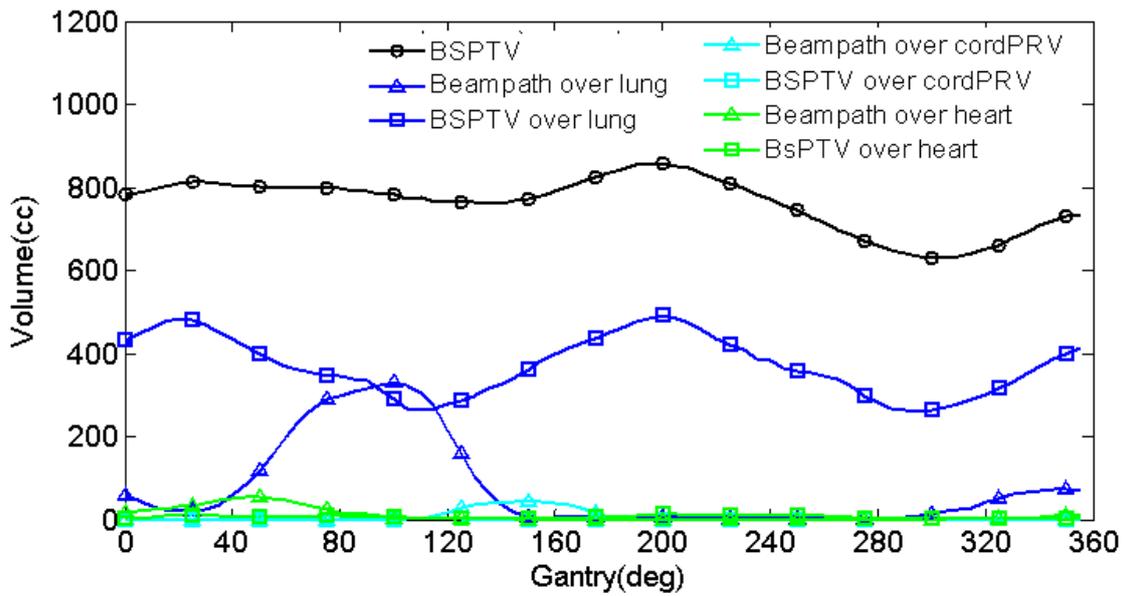

(d)

Figure 2: (a) Dose distribution and (b) DVHs for both DS (top, dashed lines) plans and PBS (bottom, solid lines) are displayed for a representative patient with locally advanced non-small cell lung cancer.

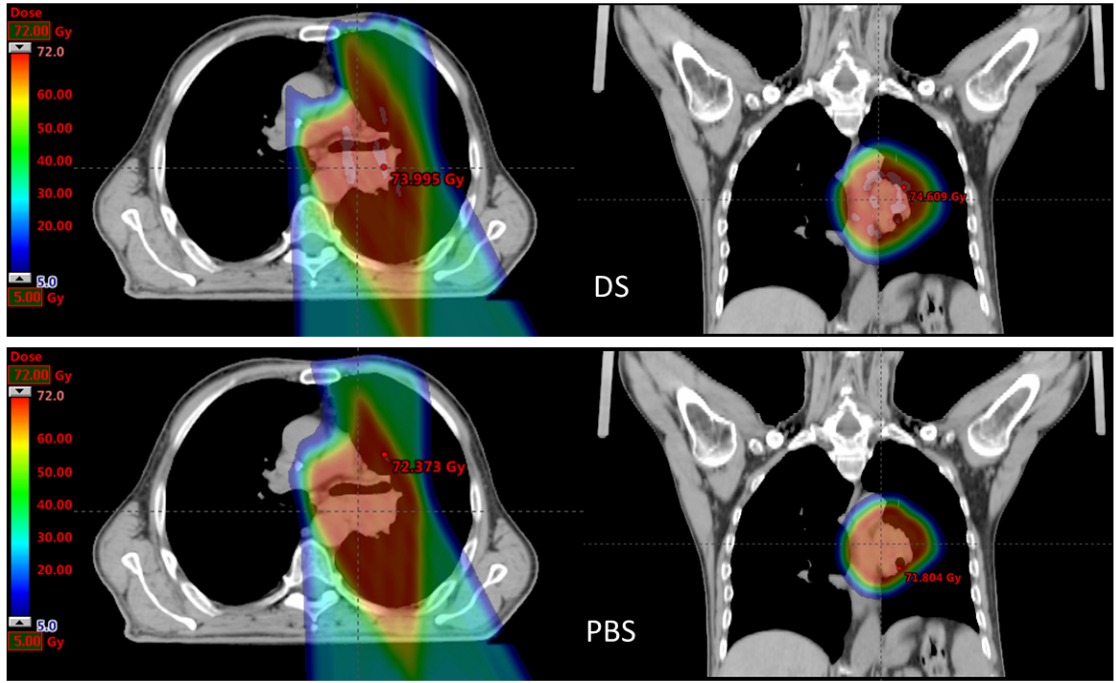

(a)

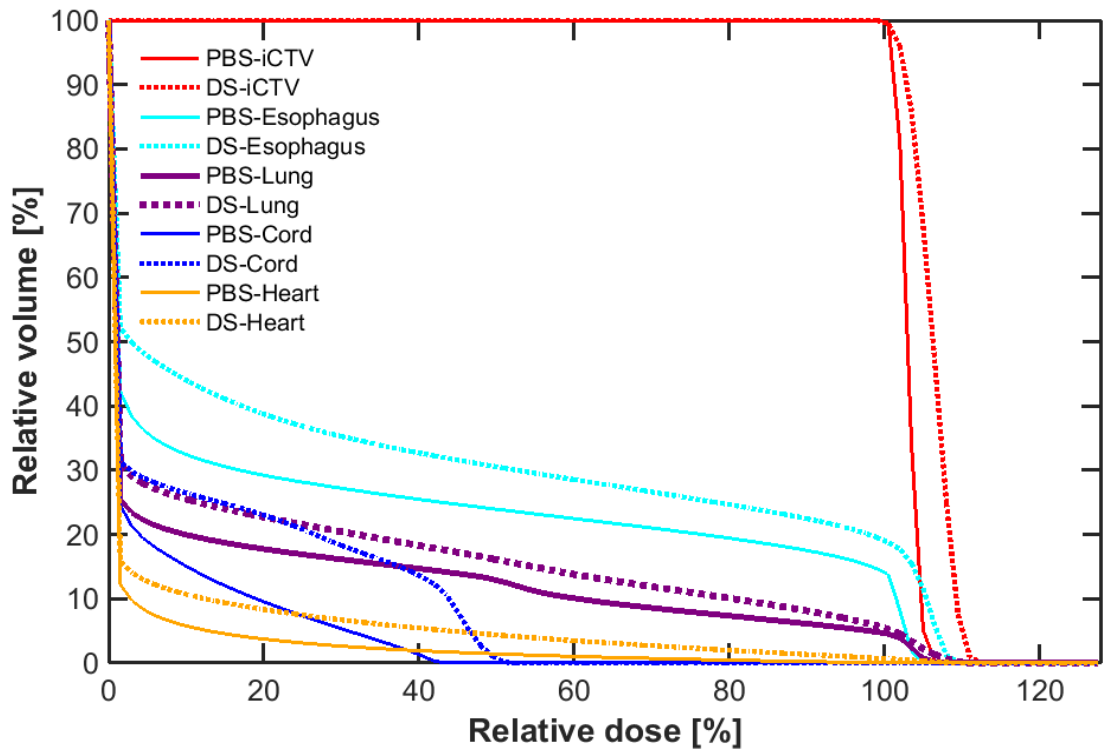

(b)

Figure 3: Average (solid lines) and 25th and 75th percentile (dashed lines) DVHs for PBS (red lines) and DS (blue lines) of (a) OAR and (b) iCTV of 10 patients with uncertainties of 3 mm setup and 3% stopping power ratio (SPR).

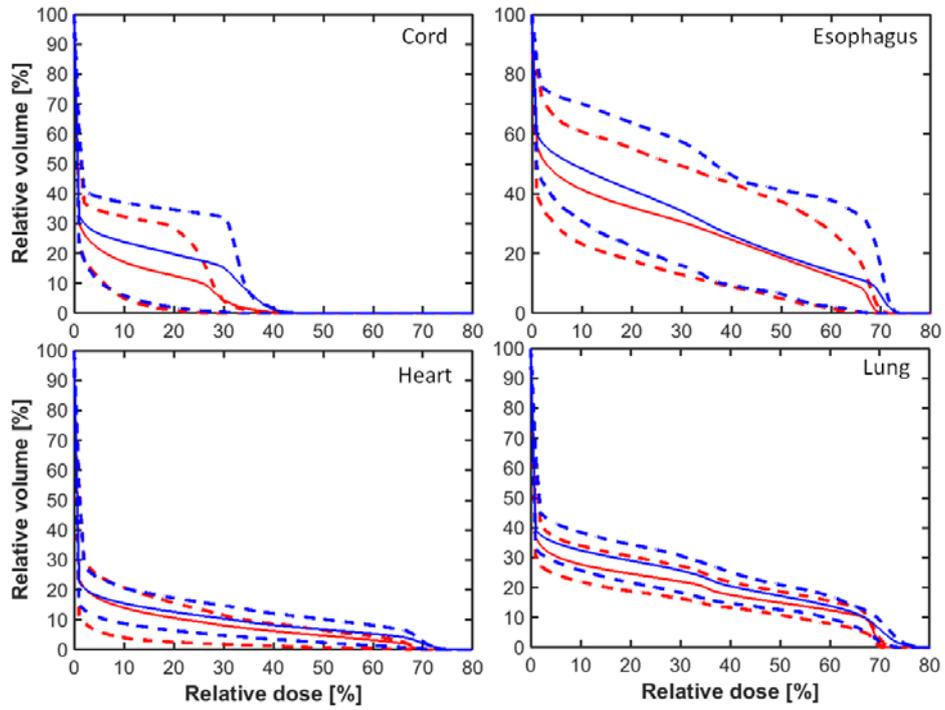

(a)

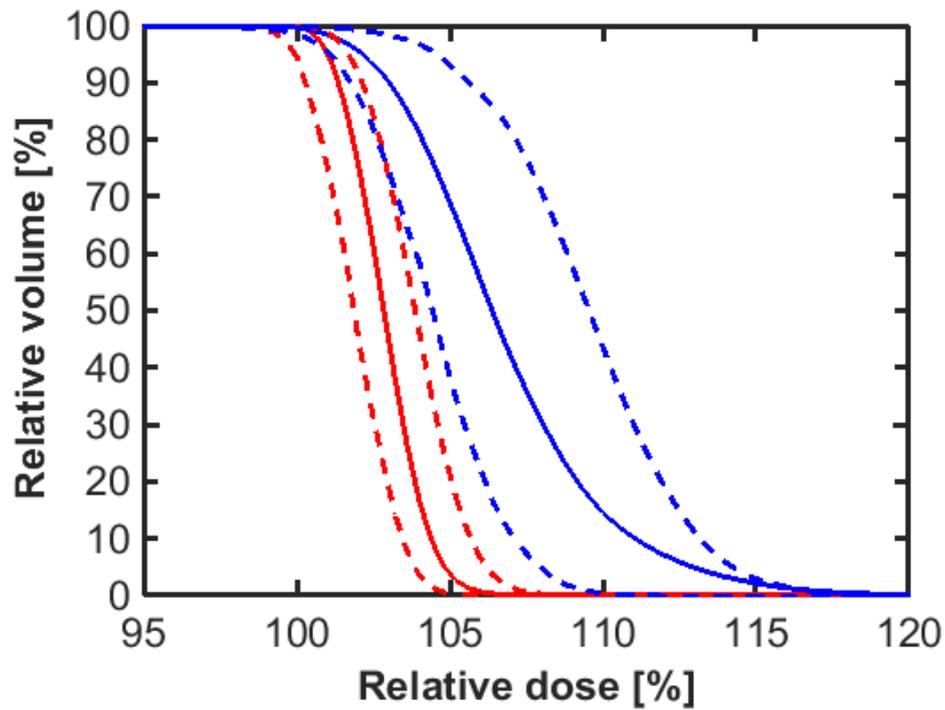

(b)

Figure 4: Beam interplay with organ motion for a representative patient with locally advanced non-small cell lung cancer. Dose distributions of one, four and eight treatment fractions are shown in top left, top right and bottom left, respectively. The bottom right shows the DVH of iCTV for one, four and eight treatment fractions accounting for interplay and original treatment plan without accounting interplay. Dark red, red, yellow, magenta and green contours represent the isodose lines of 110%, 105%, 100%, 95% and 50% of the prescription dose.

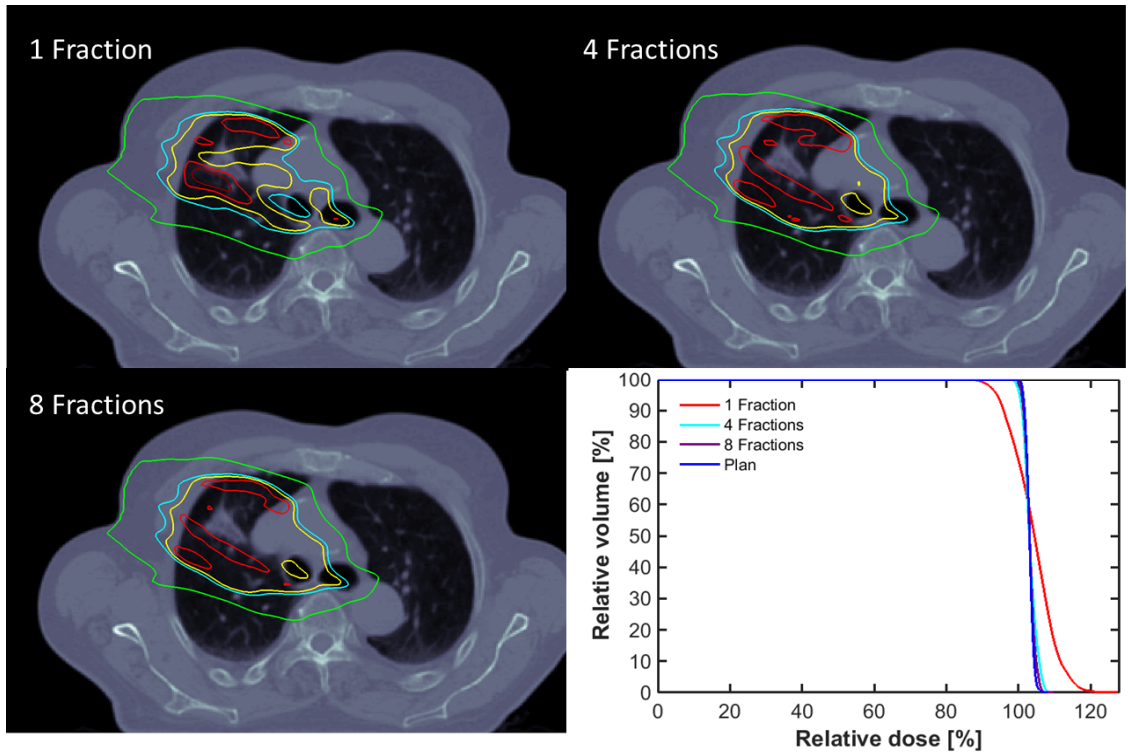

Table.1. (a) Magnitude of motion in mm observed along the beam path for the target, rib and entrance; and (b) beam specific planning target volumes in cm$^3$ due to CT/stopping power ratio, motion, and patient setup uncertainties, the quadratic and linear summations of three uncertainties for twenty beams of ten patients.

(a)

|  | Target | Rib | Entrance |
|---|---|---|---|
|  | Mean ± SE | Mean ± SE | Mean ± SE |
| Superior-Inferior (SI) | 8.3 ± 2.2 | 1.4 ± 1.2 | 0.7 ± 1.6 |
| Anterior-Posterior (AP) | 3.7 ± 1.2 | 1.3 ± 1.0 | 0.8 ± 1.7 |
| Right-Left (RL) | 2.4 ± 1.0 | 1.2 ± 1.0 | 0.5 ± 1.0 |

(b)

|  | iCTV | CT | Motion | Setup | Quadratic S | Linear S |
|---|---|---|---|---|---|---|
|  | Mean ± SE | Mean ± SE | Mean ± SE | Mean ± SE | Mean ± SE | Mean ± SE |
| Volume | 243 ± 131 | 506 ± 185 | 633 ± 203 | 482 ± 186 | 656 ± 207 | 755 ± 218 |

Table 2: Comparison of OAR dose-volume criteria, OAR maximum dose, and target coverage for PBS and DS plans in ten patients.

|  | DS | | PBS | | Difference (DS-PBS) | | |
|---|---|---|---|---|---|---|---|
|  | Mean | [25$^{th}$~75$^{th}$] | Mean | [25$^{th}$~75$^{th}$] | Mean | [25$^{th}$~75$^{th}$] | p-value |
| Lung | | | | | | | |
| V5 (%) | 35.0 | [28.7~41.4] | 30.7 | [24.4~37.0] | 4.3 | [2.8~5.5] | 0.003 |
| V20 (%) | 29.0 | [21.7~34.4] | 24.6 | [18.8~30.7] | 4.4 | [1.9~6.6] | 0.001 |
| Mean dose (Gy) | 16.4 | [11.7~19.7] | 14.1 | [10.5~17.2] | 2.3 | [1.4~3.2] | <0.001 |
| Heart | | | | | | | |
| V30 (%) | 10.5 | [4.8~14.8] | 8.2 | [1.8~11.5] | 2.3 | [1.1~3.3] | 0.004 |
| V45 (%) | 7.5 | [2.9~11.2] | 5.5 | [0.9~7.7] | 2.0 | [0.9~3.4] | 0.146 |
| Max dose | | | | | | | |
| iCTV (Gy) | 77.2 | [75.1~79.1] | 72.3 | [71.6~72.9] | 5.0 | [3.1~6.5] | <0.001 |
| Cord (Gy) | 37.2 | [34.8~45.4] | 31.9 | [28.0~42.2] | 5.3 | [1.2~7.3] | 0.012 |
| Esophagus (Gy) | 71.6 | [69.3~73.4] | 67.8 | [69.4~71.1] | 3.7 | [0.7~4.1] | 0.049 |
| Heart (Gy) | 69.1 | [71.7~75.4] | 64.7 | [70.2~71.3] | 4.5 | [1.5~5.0] | 0.016 |
| Criteria | | | | | | | |
| ID(GyL) | 1.30 | [0.87~1.67] | 1.10 | [0.78~1.32] | 0.2 | [0.08~0.34] | 0.002 |
| CI | 3.2 | [2.3~3.7] | 2.8 | [2.1~3.4] | 0.4 | [-0.06~0.75] | 0.056 |
| CQ | 0.99 | [0.99~1.00] | 1.00 | [1.00~1.00] | -0.01 | [-0.01~0.00] | 0.007 |
| HI | 0.08 | [0.06~0.09] | 0.04 | [0.04~0.04] | 0.04 | [0.03~0.05] | <0.001 |